\pdfoutput=1
\documentclass[12pt,a4paper]{article}

\usepackage{ifthen} 
\newboolean{pdflatex}
\setboolean{pdflatex}{true} 

\newboolean{articletitles}
\setboolean{articletitles}{true} 

\newboolean{uprightparticles}
\setboolean{uprightparticles}{false} 

\def\paperauthors{LHCb collaboration} 
\def\paperasciititle{Template for writing LHCb papers} 
\def\papertitle{Measurement of \CP asymmetries\\ in \DorDstoEtaPi and \DorDstoEtapPi decays} 
\def\paperkeywords{{High Energy Physics}, {LHCb}} 
\def\papercopyright{\the\year\ CERN for the benefit of the LHCb collaboration} 
\def\paperlicence{CC BY 4.0 licence}
\def\paperlicenceurl{https://creativecommons.org/licenses/by/4.0/}

\usepackage[top=1in, bottom=1.25in, left=1in, right=1in]{geometry}

\usepackage{microtype}
\usepackage{lineno}  
\usepackage{xspace} 
\usepackage{caption} 

\usepackage{graphicx}  
\usepackage{color}
\usepackage{colortbl}
\graphicspath{{./figs/}} 

\usepackage{amsmath} 
\usepackage{amssymb}
\usepackage{amsfonts}
\usepackage{upgreek} 

\newcommand*\patchAmsMathEnvironmentForLineno[1]{%
\expandafter\let\csname old#1\expandafter\endcsname\csname #1\endcsname
\expandafter\let\csname oldend#1\expandafter\endcsname\csname
end#1\endcsname
 \renewenvironment{#1}%
   {\linenomath\csname old#1\endcsname}%
   {\csname oldend#1\endcsname\endlinenomath}%
}
\newcommand*\patchBothAmsMathEnvironmentsForLineno[1]{%
  \patchAmsMathEnvironmentForLineno{#1}%
  \patchAmsMathEnvironmentForLineno{#1*}%
}
\AtBeginDocument{%
\patchBothAmsMathEnvironmentsForLineno{equation}%
\patchBothAmsMathEnvironmentsForLineno{align}%
\patchBothAmsMathEnvironmentsForLineno{flalign}%
\patchBothAmsMathEnvironmentsForLineno{alignat}%
\patchBothAmsMathEnvironmentsForLineno{gather}%
\patchBothAmsMathEnvironmentsForLineno{multline}%
\patchBothAmsMathEnvironmentsForLineno{eqnarray}%
}

\usepackage{hyperxmp}

\usepackage[pdftex,
            pdfauthor={\paperauthors},
            pdftitle={\paperasciititle},
            pdfkeywords={\paperkeywords},
            pdfcopyright={Copyright (C) \papercopyright},
            pdflicenseurl={\paperlicenceurl}]{hyperref}

\usepackage[colorinlistoftodos,textsize=scriptsize]{todonotes}

\usepackage[bottom,flushmargin,hang,multiple]{footmisc}

\usepackage[all]{hypcap} 

\usepackage{xspace} 
\usepackage{upgreek}

\def\lhcb   {\mbox{LHCb}\xspace}

\def\MagUp {\mbox{\em Mag\kern -0.05em Up}\xspace}

\ifthenelse{\boolean{uprightparticles}}%
{

 \def\Peta        {\ensuremath{\upeta}\xspace}

 \def\Ppi         {\ensuremath{\uppi}\xspace}

 \def\Pphi        {\ensuremath{\upphi}\xspace}

 \def\PDelta      {\ensuremath{\Delta}\xspace}                 
 \def\PXi         {\ensuremath{\Xi}\xspace}                 
 \def\PLambda     {\ensuremath{\Lambda}\xspace}                 
 \def\PSigma      {\ensuremath{\Sigma}\xspace}                 
 \def\POmega      {\ensuremath{\Omega}\xspace}                 
 \def\PUpsilon    {\ensuremath{\Upsilon}\xspace}
 \let\oldPi\Pi
 \def\PPi         {\ensuremath{\oldPi}\xspace}

 \def\PB      {\ensuremath{\mathrm{B}}\xspace}                 
                  
 \def\PD      {\ensuremath{\mathrm{D}}\xspace}

 \def\PK      {\ensuremath{\mathrm{K}}\xspace}

 \def\Pb      {\ensuremath{\mathrm{b}}\xspace}                 
 \def\Pc      {\ensuremath{\mathrm{c}}\xspace}

 \def\Pi      {\ensuremath{\mathrm{i}}\xspace}

 \def\Ps      {\ensuremath{\mathrm{s}}\xspace}

 \def\thebaroffset{0.0em}
}
{

 \def\Peta        {\ensuremath{\eta}\xspace}

 \def\Ppi         {\ensuremath{\pi}\xspace}

 \def\Pphi        {\ensuremath{\phi}\xspace}

 \mathchardef\PDelta="7101
 \mathchardef\PXi="7104
 \mathchardef\PLambda="7103
 \mathchardef\PSigma="7106
 \mathchardef\POmega="710A
 \mathchardef\PUpsilon="7107
 \mathchardef\PPi="7105
                  
 \def\PB      {\ensuremath{B}\xspace}                 
                  
 \def\PD      {\ensuremath{D}\xspace}

 \def\PK      {\ensuremath{K}\xspace}

 \def\Pb      {\ensuremath{b}\xspace}                 
 \def\Pc      {\ensuremath{c}\xspace}

 \def\Pi      {\ensuremath{i}\xspace}

 \def\Ps      {\ensuremath{s}\xspace}

 \def\thebaroffset{0.18em}
}
\newcommand{\offsetoverline}[2][\thebaroffset]{\kern #1\overline{\kern -#1 #2}}%

\makeatletter
\ifcase \@ptsize \relax
  \newcommand{\miniscule}{\@setfontsize\miniscule{4}{5}}
\or
  \newcommand{\miniscule}{\@setfontsize\miniscule{5}{6}}
\or
  \newcommand{\miniscule}{\@setfontsize\miniscule{5}{6}}
\fi
\makeatother

\DeclareRobustCommand{\optbar}[1]{\shortstack{{\miniscule (\rule[.5ex]{1.25em}{.18mm})}
  \\ [-.7ex] $#1$}}

\def\squark    {{\ensuremath{\Ps}}\xspace}

\def\cquark    {{\ensuremath{\Pc}}\xspace}

\def\bquark    {{\ensuremath{\Pb}}\xspace}

\def\pion   {{\ensuremath{\Ppi}}\xspace}
\def\piz    {{\ensuremath{\pion^0}}\xspace}
\def\pip    {{\ensuremath{\pion^+}}\xspace}
\def\pim    {{\ensuremath{\pion^-}}\xspace}
\def\pipm   {{\ensuremath{\pion^\pm}}\xspace}

\def\kaon    {{\ensuremath{\PK}}\xspace}

\def\KorKbar {\kern \thebaroffset\optbar{\kern -\thebaroffset \PK}{}\xspace}

\def\Kp      {{\ensuremath{\kaon^+}}\xspace}
\def\Km      {{\ensuremath{\kaon^-}}\xspace}

\newcommand{\etaz}{\ensuremath{\Peta}\xspace}
\newcommand{\etapr}{\ensuremath{\Peta^{\prime}}\xspace}
\newcommand{\phiz}{\ensuremath{\Pphi}\xspace}

\def\D       {{\ensuremath{\PD}}\xspace}

\def\DorDbar {\kern \thebaroffset\optbar{\kern -\thebaroffset \PD}\xspace}

\def\Dp      {{\ensuremath{\D^+}}\xspace}
\def\Dm      {{\ensuremath{\D^-}}\xspace}

\def\DpDm    {\ensuremath{\Dp {\kern -0.16em \Dm}}\xspace}

\def\Ds      {{\ensuremath{\D^+_\squark}}\xspace}
\def\Dsp     {{\ensuremath{\D^+_\squark}}\xspace}

\def\B       {{\ensuremath{\PB}}\xspace}

\def\BorBbar {\kern \thebaroffset\optbar{\kern -\thebaroffset \PB}\xspace}

\def\Bd      {{\ensuremath{\B^0}}\xspace}

\def\BdorBdbar {\kern \thebaroffset\optbar{\kern -\thebaroffset \Bd}\xspace}

\def\Bs      {{\ensuremath{\B^0_\squark}}\xspace}

\def\BsorBsbar {\kern \thebaroffset\optbar{\kern -\thebaroffset \Bs}\xspace}

\def\Y#1S{\ensuremath{\PUpsilon{(#1S)}}\xspace}

\def\Lz          {{\ensuremath{\PLambda}}\xspace}

\def\LorLbar     {\kern \thebaroffset\optbar{\kern -\thebaroffset \PLambda}\xspace}

\def\Lc          {{\ensuremath{\Lz^+_\cquark}}\xspace}

\def\to                 {\ensuremath{\rightarrow}\xspace}

\def\CP                {{\ensuremath{C\!P}}\xspace}

\newcommand{\ACP}{{\ensuremath{{\mathcal{A}}^{\CP}}}\xspace}

\def\AT#1     {\ensuremath{A_{\mathrm{T}}^{#1}}\xspace}           

\def\C#1      {\ensuremath{\mathcal{C}_{#1}}\xspace}                       
\def\Cp#1     {\ensuremath{\mathcal{C}_{#1}^{'}}\xspace}                    
\def\Ceff#1   {\ensuremath{\mathcal{C}_{#1}^{\mathrm{(eff)}}}\xspace}        
\def\Cpeff#1  {\ensuremath{\mathcal{C}_{#1}^{'\mathrm{(eff)}}}\xspace}       
\def\Ope#1    {\ensuremath{\mathcal{O}_{#1}}\xspace}                       
\def\Opep#1   {\ensuremath{\mathcal{O}_{#1}^{'}}\xspace}                    

\newcommand{\nospaceunit}[1]{\ensuremath{\text{#1}}}       
\newcommand{\aunit}[1]{\ensuremath{\text{\,#1}}}       

\newcommand{\tev}{\aunit{Te\kern -0.1em V}\xspace}
\newcommand{\gev}{\aunit{Ge\kern -0.1em V}\xspace}
\newcommand{\mev}{\aunit{Me\kern -0.1em V}\xspace}
\newcommand{\kev}{\aunit{ke\kern -0.1em V}\xspace}
\newcommand{\ev}{\aunit{e\kern -0.1em V}\xspace}
 
\newcommand{\mevc}{\ensuremath{\aunit{Me\kern -0.1em V\!/}c}\xspace}
\newcommand{\gevc}{\ensuremath{\aunit{Ge\kern -0.1em V\!/}c}\xspace}
\newcommand{\mevcc}{\ensuremath{\aunit{Me\kern -0.1em V\!/}c^2}\xspace}
\newcommand{\gevcc}{\ensuremath{\aunit{Ge\kern -0.1em V\!/}c^2}\xspace}

\def\mum  {\ensuremath{\,\upmu\nospaceunit{m}}\xspace}

\def\fb   {\ensuremath{\aunit{fb}}\xspace}
\def\invfb   {\ensuremath{\fb^{-1}}\xspace}

\newcommand{\chisq}{\ensuremath{\chi^2}\xspace}

\def\gsim{{~\raise.15em\hbox{$>$}\kern-.85em
          \lower.35em\hbox{$\sim$}~}\xspace}
\def\lsim{{~\raise.15em\hbox{$<$}\kern-.85em
          \lower.35em\hbox{$\sim$}~}\xspace}

\def\pt         {\ensuremath{p_{\mathrm{T}}}\xspace}

\def\ptot       {\ensuremath{p}\xspace}

\def\evtgen     {\mbox{\textsc{EvtGen}}\xspace}

\def\geant      {\mbox{\textsc{Geant4}}\xspace}

\def\photos     {\mbox{\textsc{Photos}}\xspace}

\def\pythia     {\mbox{\textsc{Pythia}}\xspace}

\def\tell1  {TELL1\xspace}
\def\ukl1   {UKL1\xspace}

\newcommand{\eg}{\mbox{\itshape e.g.}\xspace}
\newcommand{\ie}{\mbox{\itshape i.e.}\xspace}

\newcommand{\lhcborcid}[1]{\href{https://orcid.org/#1}{\hspace*{0.1em}\raisebox{-0.45ex}{\includegraphics[width=1em]{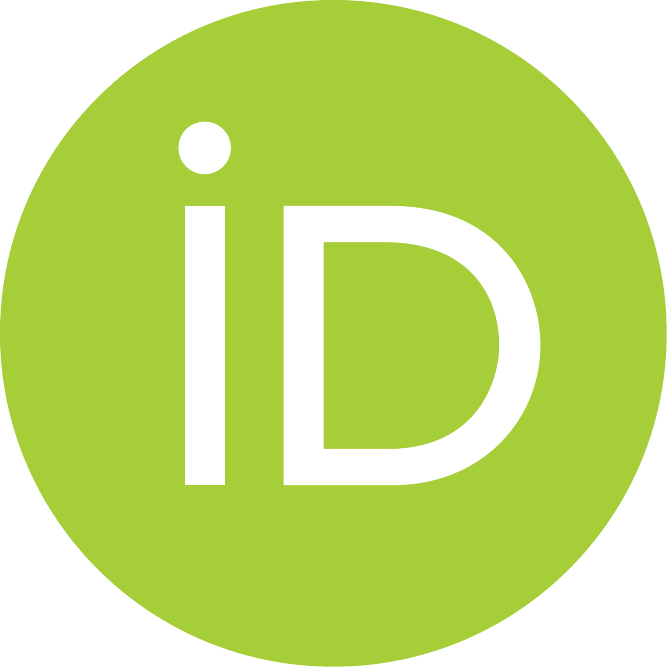}}}}

\usepackage{cite} 
\usepackage{mciteplus}

\newcommand{\etazpr}{\ensuremath{\Peta^{(\prime)}}\xspace}
\def\DorDs      {{\ensuremath{\D^+_{(\squark)}}}\xspace}
\def\DorDsm      {{\ensuremath{\D^-_{(\squark)}}}\xspace}
\def\DorDspm      {{\ensuremath{\D^{\pm}_{(\squark)}}}\xspace}

\newcommand{\DstoEtapPi}{\ensuremath{\Dsp \to \etapr \pip}\xspace}
\newcommand{\DtoEtapPi}{\ensuremath{\Dp \to \etapr \pip}\xspace}
\newcommand{\DorDstoEtapPi}{\ensuremath{\DorDs \to \etapr \pip}\xspace}

\newcommand{\DstoEtaPi}{\ensuremath{\Dsp \to \etaz \pip}\xspace}
\newcommand{\DtoEtaPi}{\ensuremath{\Dp \to \etaz \pip}\xspace}
\newcommand{\DorDstoEtaPi}{\ensuremath{\DorDs \to \etaz \pip}\xspace}

\newcommand{\DstoEtazpPi}{\ensuremath{\Dsp \to \etazpr \pip}\xspace}
\newcommand{\DtoEtazpPi}{\ensuremath{\Dp \to \etazpr \pip}\xspace}
\newcommand{\DorDstoEtazpPi}{\ensuremath{\DorDs \to \etazpr \pip}\xspace}

\newcommand{\DorDstoPhiPiThreePi}{\ensuremath{\DorDs \to \phiz_{3\pi} \pip}\xspace}

\newcommand{\etazprtogpp}{\ensuremath{\etazpr\to\gamma\pip\pim}\xspace}
\newcommand{\DstoPhiPi}{\ensuremath{\Ds \to \phiz \pip}\xspace}
\newcommand{\DtoPhiPi}{\ensuremath{\Dp \to \phiz \pip}\xspace}
\newcommand{\DorDstoPhiPi}{\ensuremath{\DorDs \to \phiz \pip}\xspace}

\newcommand{\mppg}{\ensuremath{m(\gamma\pip\pim)}\xspace}
\newcommand{\mgpp}{\ensuremath{m(\gamma\pip\pim)}\xspace}

\newcommand{\Araw}{{\ensuremath{{\mathcal{A}}^{\mathrm{ raw}}}}\xspace}
\newcommand{\Adet}{{\ensuremath{{\mathcal{A}}^{\mathrm{ det}}}}\xspace}
\newcommand{\Aprod}{{\ensuremath{{\mathcal{A}}^{\mathrm{ prod}}}}\xspace}
\newcommand{\AKK}{{\ensuremath{{\mathcal{A}}^{\mathrm{ KK}}}}\xspace}

\usepackage{longtable} 

\begin{document}

\renewcommand{\thefootnote}{\fnsymbol{footnote}}
\setcounter{footnote}{1}


\begin{titlepage}
\pagenumbering{roman}

\vspace*{-1.5cm}
\centerline{\large EUROPEAN ORGANIZATION FOR NUCLEAR RESEARCH (CERN)}
\vspace*{1.5cm}
\noindent
\begin{tabular*}{\linewidth}{lc@{\extracolsep{\fill}}r@{\extracolsep{0pt}}}
\ifthenelse{\boolean{pdflatex}}
{\vspace*{-1.5cm}\mbox{\!\!\!\includegraphics[width=.14\textwidth]{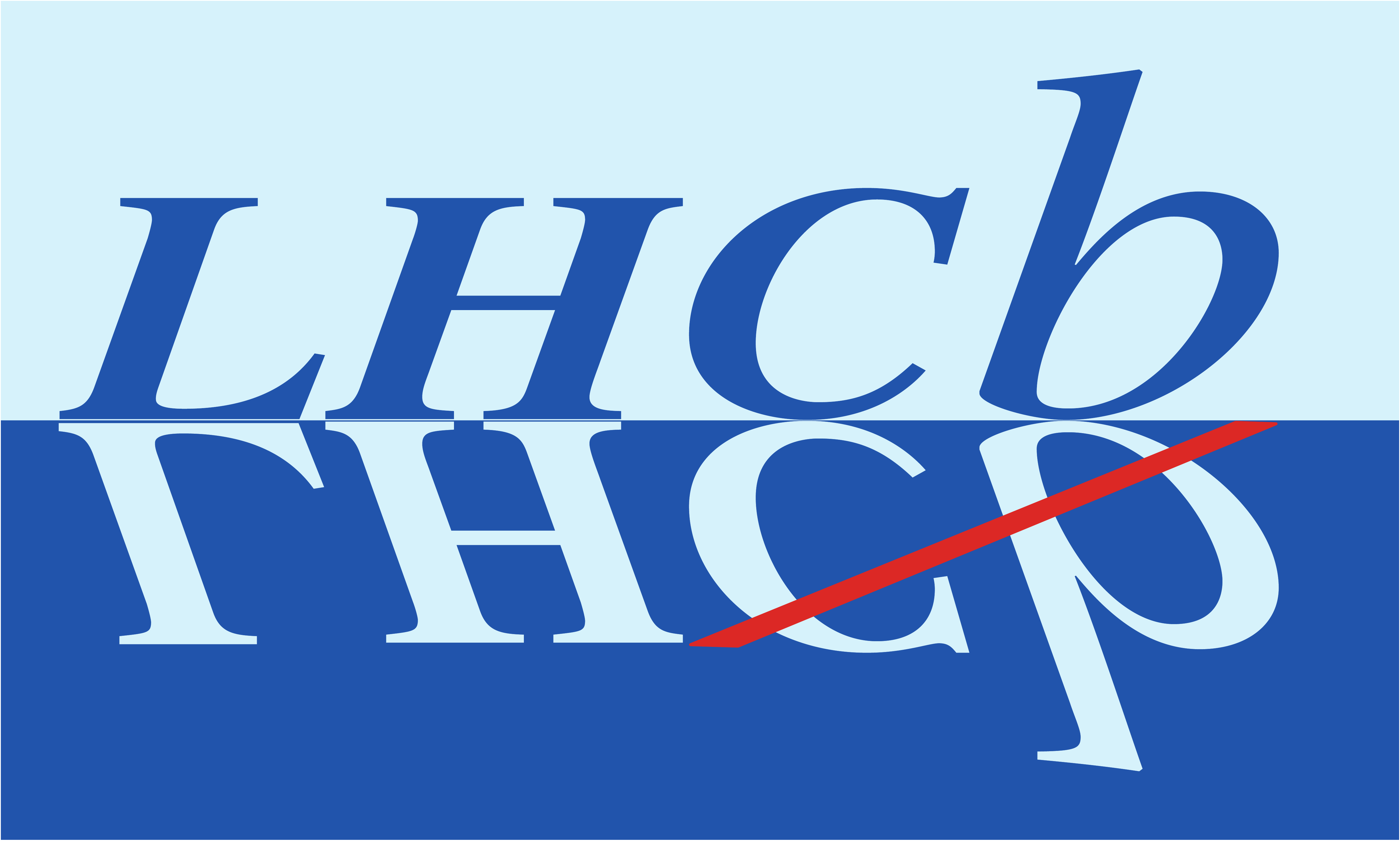}} & &}%
{\vspace*{-1.2cm}\mbox{\!\!\!\includegraphics[width=.12\textwidth]{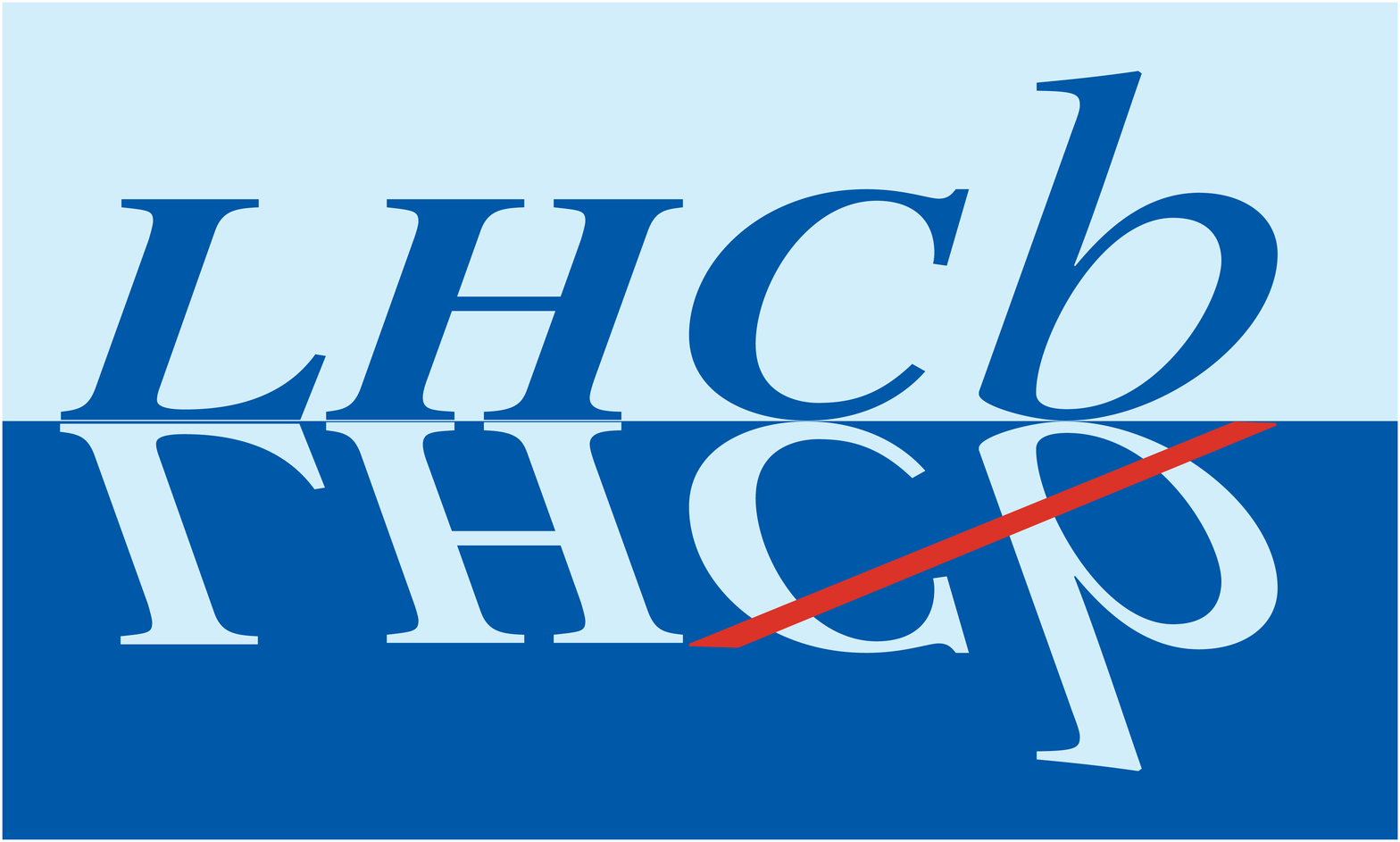}} & &}%
\\
 & & CERN-EP-2022-035 \\  
 & & LHCb-PAPER-2021-051 \\  
 & & April 26, 2023 \\ 
& & \\

\end{tabular*}

\vspace*{2.0cm} 

{\normalfont\bfseries\boldmath\huge
\begin{center}
  \papertitle 
\end{center}
}

\vspace*{1.0cm} 

\begin{center}
\paperauthors
\footnote{Authors are listed at the end of this paper.}

\end{center}

\vspace{\fill}

\begin{abstract}
  \noindent
  Searches for \CP violation in the decays \DorDstoEtaPi and \DorDstoEtapPi are performed using $pp$ collision data corresponding to 6 \invfb of integrated luminosity collected by the \lhcb experiment. The calibration channels \DorDstoPhiPi are used to remove production and detection asymmetries. The resulting \CP-violating asymmetries are
  \begin{align*}
      \ACP(\DtoEtaPi)   & = (0.34 \pm 0.66 \pm 0.16 \pm 0.05)\%, \\
      \ACP(\DstoEtaPi)  & = (0.32 \pm 0.51 \pm 0.12)\%, \\
      \ACP(\DtoEtapPi)  & = (0.49 \pm 0.18 \pm 0.06 \pm 0.05)\%, \\
      \ACP(\DstoEtapPi) & = (0.01 \pm 0.12 \pm 0.08)\%, 
  \end{align*}
  where the first uncertainty is statistical, the second is  systematic and the third, relevant for the \Dp channels, is due to the uncertainty on $\ACP(\Dp\to\phi\pip)$. These measurements, currently the most precise for three of the four channels considered, are consistent with the absence of \CP violation. A combination of these results with previous \lhcb measurements is presented.

\end{abstract}

\vspace*{2.0cm}

\begin{center}
  Published in JHEP 04 (2023) 081
\end{center}

\vspace{\fill}

{\footnotesize 
\centerline{\copyright~\papercopyright. \href{\paperlicenceurl}{\paperlicence}.}}
\vspace*{2mm}

\end{titlepage}


\newpage
\setcounter{page}{2}
\mbox{~}
%
%
%
%


\renewcommand{\thefootnote}{\arabic{footnote}}
\setcounter{footnote}{0}

\cleardoublepage


\pagestyle{plain} 
\setcounter{page}{1}
\pagenumbering{arabic}




\section{Introduction}
\label{sec:Introduction}
The observed baryon asymmetry in the Universe requires the violation of the charge-conjugation parity (\CP) symmetry~\cite{Sakharov:1967dj}. Within the Standard Model (SM), \CP violation occurs due to an irreducible complex phase in the Cabibbo-Kobayashi-Maskawa (CKM) matrix, which describes the transitions between quarks~\cite{Cabibbo:1963yz,Kobayashi:1973fv}. However, the size of \CP violation in the SM appears to be too small to explain the observed baryon asymmetry, suggesting that there may be additional sources of \CP violation beyond the SM~\cite{Cohen:1993nk,Riotto:1999yt,Hou:2008xd}. Such sources can be associated to new heavy particles, giving further motivation for \CP violation searches.

Although \CP violation in the $b$- and $s$-quark sectors has been established for some time~\cite{BaBar:2004gyj,Belle:2004nch,LHCb-PAPER-2013-018,LHCb-PAPER-2012-001}, its observation in the charm sector was only accomplished in 2019~\cite{LHCb-PAPER-2019-006}. The charm sector provides a unique opportunity to study \CP violation in decays of up-type quarks. Charge-parity violation in charm decays occurs through the interference of tree- and loop-level diagrams in the Cabibbo-suppressed quark transitions $c\to\bar ddu$ and $c\to \bar ssu$. Standard Model calculations of \CP violation in the charm sector are difficult due to the presence of low-energy strong-interaction effects; as a consequence, predictions of \CP asymmetries vary by orders of magnitude, typically bewteen $10^{-4}$ and $10^{-2}$~\cite{Grossman:2006jg,Golden:1989qx,Buccella:1994nf,Bianco:2003vb,Artuso:2008vf}. 
The study of \CP asymmetries in 
several decays related by flavour symmetry can provide insight to the origin of observed \CP violation~\cite{Pirtskhalava:2011va,Cheng:2012wr,Feldmann:2012js,Li:2012cfa,Franco:2012ck,Brod:2012ud,Atwood:2012ac,Hiller:2012xm,Muller:2015rna}. 
Of particular interest are two-body decays of \Dp and \Ds mesons, such as 
\DstoEtazpPi (Cabibbo-favoured) and  \DtoEtazpPi (singly Cabibbo-suppressed).\footnote{Charge-conjugate decays are implied throughout this article, except when discussing asymmetries.} For the singly Cabibbo suppressed decays, predictions for the \CP asymmetry from topological-diagram~\cite{Cheng:2012wr} and factorization~\cite{Li:2012cfa} approaches are, respectively, 
$-0.65\times 10^{-3}$
and $-0.26\times 10^{-3}$
for the \DtoEtaPi decay mode. 
For the \DtoEtapPi channel the corresponding predictions are $0.41\times 10^{-3}$ and $1.18\times 10^{-3}$. The \CP asymmetries for these channels have been measured by the CLEO~\cite{Mendez:2009aa,Onyisi:2013bjt}, Belle~\cite{Won:2011ku,Belle:2021ygw} and LHCb~\cite{LHCb-PAPER-2016-041,LHCb-PAPER-2021-001} collaborations. No evidence for \CP violation has been observed, within the uncertainties of a few per mil. 

This article presents measurements of \CP asymmetries for the 
modes \DorDstoEtaPi and \DorDstoEtapPi
at the LHCb experiment, made using proton-proton ($pp$) collisions, recorded during the period 2015--2018 at a centre-of-mass energy $\sqrt{s}=13\tev$, corresponding to an integrated luminosity of 6 \invfb. The \etaz and \etapr mesons are both reconstructed in the final state $\gamma\pip\pim$. The two charged particles in the final states allow the reconstruction of the \etazpr decay vertex.

The \CP asymmetry is defined as
\begin{equation}
    \ACP(\DorDs\to f^+) \equiv  \frac{\Gamma(\DorDs\to f^+) - \Gamma(\DorDsm\to f^-)}{\Gamma(\DorDs\to f^+) + \Gamma(\DorDsm\to f^-)},
\end{equation}
where $f$ is the considered final state and $\Gamma$ is the partial decay width.
Experimentally, the raw asymmetry \Araw is measured using the event yields $N$ as
\begin{equation}
    \Araw(\DorDs\to f^+) \equiv  \frac{N(\DorDs\to f^+) - N(\DorDsm\to f^-)}{N(\DorDs\to f^+) + N(\DorDsm\to f^-)}.
\end{equation}
The difference between \ACP and \Araw arises from asymmetries in the production of positively and negatively charged $D_{(s)}^{\pm}$ mesons~\cite{LHCb-PAPER-2018-010,LHCb-PAPER-2012-026} and in the detection efficiency of the corresponding final states. For small asymmetries, \Araw may be approximated to first order as
\begin{equation}
    \Araw(\DorDs\to f^+) \approx \ACP(\DorDs\to f^+) + \Aprod(\DorDs) + \Adet(f^+),
\end{equation}
where $\Aprod = \frac{\sigma(\DorDs)-\sigma(\DorDsm)}{\sigma(\DorDs)+\sigma(\DorDsm)}$ is the production asymmetry and $\Adet = \frac{\epsilon(f^+)-\epsilon(f^-)}{\epsilon(f^+)+\epsilon(f^-)}$ is the detection asymmetry. Here $\sigma(\DorDspm)$ is the production cross section for \DorDspm mesons and $\epsilon(f^{\pm})$ is the efficiency for detecting the final state $f^{\pm}$. 
The detection asymmetry can arise due to instrumental effects, such as different interaction cross-sections of positive and negative particles with the detector material, or a small charge-dependence of geometrical effects that are not fully treated in the reconstruction algorithms. For $f^{\pm} = \etazpr\pipm$, \Adet is due only to the pion, since the decays $\etazpr\to\gamma\pip\pim$ are charge symmetric. 

The production and detection asymmetries are subtracted using the control channels \DstoPhiPi and \DtoPhiPi, followed by the decay $\phiz\to\Kp\Km$. The production and detection asymmetries are the same for the control channels and the signal channels, after accounting for small differences in the kinematic properties, 
and are therefore eliminated in the difference of the raw asymmetries
\begin{align}
    \Araw(\DtoEtazpPi) - \Araw(\DtoPhiPi) &= \ACP(\DtoEtazpPi) - \ACP(\DtoPhiPi), \label{eq:araw1}\\
    \Araw(\DstoEtazpPi) - \Araw(\DstoPhiPi) &= \ACP(\DstoEtazpPi). \label{eq:araw2}
\end{align}
The Cabibbo-favoured channel \DstoPhiPi is assumed to have \mbox{$\ACP=0$}, based on theoretical SM  expectations~\cite{Bianco:2003vb}.
For the singly Cabibbo-suppressed control channel, the value of \mbox{$\ACP(\DtoPhiPi)= (0.005 \pm 0.051)\%$~\cite{LHCb-PAPER-2019-002}} is taken as an external input. Its uncertainty is small compared to the sensitivity of this measurement. 

\section{Detector and simulation}
\label{sec:Detector}

The \lhcb detector~\cite{LHCb-DP-2008-001,LHCb-DP-2014-002} is a single-arm forward
spectrometer covering the \mbox{pseudorapidity} range $2<\eta <5$,
designed for the study of particles containing \bquark or \cquark
quarks. The detector includes a high-precision tracking system
consisting of a silicon-strip vertex detector surrounding the $pp$
interaction region, a large-area silicon-strip detector located
upstream of a dipole magnet with a bending power of about
$4{\mathrm{\,Tm}}$, and three stations of silicon-strip detectors and straw
drift tubes
placed downstream of the magnet.
The tracking system provides a measurement of the momentum, \ptot, of charged particles with
a relative uncertainty that varies from 0.5\% at low momentum to 1.0\% at 200\gev.\footnote{Natural units, with $c=\hbar=1$, are used throughout.}
The minimum distance of a track to a primary $pp$ collision vertex (PV), the impact parameter (IP), 
is measured with a resolution of $(15+29/\pt)\mum$,
where \pt is the component of the momentum transverse to the beam, in\,\gev.
Different types of charged hadrons are distinguished using information
from two ring-imaging Cherenkov detectors. 
Photons, electrons and hadrons are identified by a calorimeter system consisting of
scintillating-pad and preshower detectors, an electromagnetic
and a hadronic calorimeter. Muons are identified by a
system composed of alternating layers of iron and multiwire
proportional chambers.
The online event selection is performed by a trigger, 
which consists of a hardware stage, based on information from the calorimeter and muon
systems, followed by two software stages, which apply a full event
reconstruction. Detector calibration and alignment is performed 
after the first software stage and is used in the second stage.

Simulation is required to model the effects of the detector acceptance and the imposed selection requirements, and to study background contributions.
  In the simulation, $pp$ collisions are generated using
  \pythia~\cite{Sjostrand:2007gs,*Sjostrand:2006za} 
  with a specific \lhcb configuration~\cite{LHCb-PROC-2010-056}.
  Decays of unstable particles
  are described by \evtgen~\cite{Lange:2001uf}, in which final-state
  radiation is generated using \photos~\cite{davidson2015photos}.
  The interaction of the generated particles with the detector, and its response,
  are implemented using the \geant
  toolkit~\cite{Allison:2006ve, *Agostinelli:2002hh} as described in
  Ref.~\cite{LHCb-PROC-2011-006}.

\section{Event selection}
\label{sec:selection}
The online selection of signal candidates is performed by the trigger. The hardware trigger selects candidates where one or more of the \DorDs decay products produces a significant energy deposit in the calorimeter. Events where particles that are not used to reconstruct the \DorDs candidate satisfy the hardware trigger are also accepted. The first software stage requires at least one pion of the \etazpr decay to have high transverse momentum and to be well detached from all primary $pp$ interaction vertices. In the second software stage, each selected event is required to have at least one fully reconstructed \DorDstoEtazpPi candidate.

Offline, the \etazprtogpp decays are reconstructed with two oppositely charged well reconstructed tracks and a high-quality photon candidate. A multivariate classifier combining several track-quality criteria 
is applied to reject fake tracks~\cite{LHCb-DP-2014-002}. Multivariate algorithms based on the shape parameters of the cluster in the electromagnetic calorimeter are used to distinguish the photon from non-electromagnetic
deposits associated to hadrons and to reject background in which the two photons are reconstructed as a single cluster~\cite{LHCb-DP-2020-001}.
The charged tracks must have transverse momentum $\pt > 500 \mev$ and momentum $p>1000 \mev$. Furthermore, they must have particle-identification characteristics compatible with the pion hypothesis. The pion candidates must be well-detached from all primary vertices in the event. Photon candidates must satisfy $\pt>1000 \mev$. For \etapr decays, the invariant mass of the pion pair must be greater than $600 \mev$, which selects $\rho^0$ mesons from the dominant decay process $\etapr\to\rho^0\gamma$, followed by $\rho^0\to\pip\pim$. The \etazpr candidates are combined with an additional high-quality track that is consistent with being a pion, denoted the ``companion'' pion, to form \DorDs candidates. The companion pion must satisfy $ 1 < \pt < 20 \gev$, have pseudorapidity between 2 and 5, and be significantly detached from all primary vertices. Furthermore, it is required to satisfy fiducial requirements designed to exclude momentum regions for which candidates of one charge are more likely to be deflected out of the detector acceptance, causing large detection asymmetries~\cite{LHCb-PAPER-2019-006,LHCb-PAPER-2019-002}.

The \DorDs candidates must satisfy $\pt > 2000\mev$ and have proper decay time \mbox{$\tau>0.25 \aunit{ps}$}. Furthermore, the invariant mass of the three pion candidates combined must satisfy $m_{3\pi}<1825\mev$, which removes background $\DorDs \to \pip\pim\pip$ decays. 
A kinematic fit is performed constraining the three-pion candidates to arise from a common vertex and the \DorDs candidate to originate at the primary vertex~\cite{Hulsbergen:2005pu}. 
Selected candidates are required to have a $p$-value $> 1$\% for this fit. The invariant mass of the selected candidates is calculated by repeating the kinematic fit while constraining the \etazpr particle to its known mass~\cite{PDG2020}. The symbols $m(\etazpr\pip)$ are used to denote the \DorDs candidate mass calculated using the \etazpr mass constraint.
Multiple candidates are found in 5.5\% of the events. In case of more than one candidate per event, one of the candidates is selected randomly. 

The data is divided into eight subsamples, separated by year of data taking, magnet polarity and if the event had multiple primary vertices.
The small amount of data from 2015 is combined with the 2016 data. Furthermore, for the years 2015--2017, the dedicated second level software trigger only accepted events with one primary vertex, while for 2018 this requirement was eliminated. 
The analysis is performed independently on each subsample and a weighted sum is performed to determine the result for the complete dataset. 

\section{Background sources}
\label{sec:backgrounds}

The dominant background in all the signal channels is combinatorial, in which random combinations of charged particles and photons result in masses $m(\gamma\pip\pim)$ and $m(\gamma\pip\pim\pip)$ that are compatible with those of the \etazpr and the \DorDs mesons, respectively. Included in the combinatorial background is a small contribution from correctly reconstructed \etazpr mesons combined with a random companion pion.

Secondary decays, in which a true \DorDs meson is the decay product of a \bquark hadron and not produced at the primary vertex, contribute to the signal sample. Such decays can bias the \CP asymmetry measurement, because the  asymmetry in the production and decay of the \bquark hadron may differ from that of the prompt production of \DorDs mesons. The secondary background is estimated from simulation to be 11\% for the \Dp sample and 13\% for the \Ds sample. The potential bias introduced by the secondary background is treated as a source of systematic uncertainty (Sec.~\ref{sec:systematics}).

Additionally, background arising from the decay channel $\DorDs\to\phi\pip$, followed by the decay \mbox{$\phi\to\pip\pim\piz$}, (henceforth denoted \DorDstoPhiPiThreePi) contributes for the channels with \etapr in the final state. This can occur when the \piz meson is misreconstructed as a photon. A component for this background is accounted for in the fit, as discussed in Sec.~\ref{sec:fit}. 

\begin{figure}
    \centering
    \includegraphics[width=.48\textwidth]{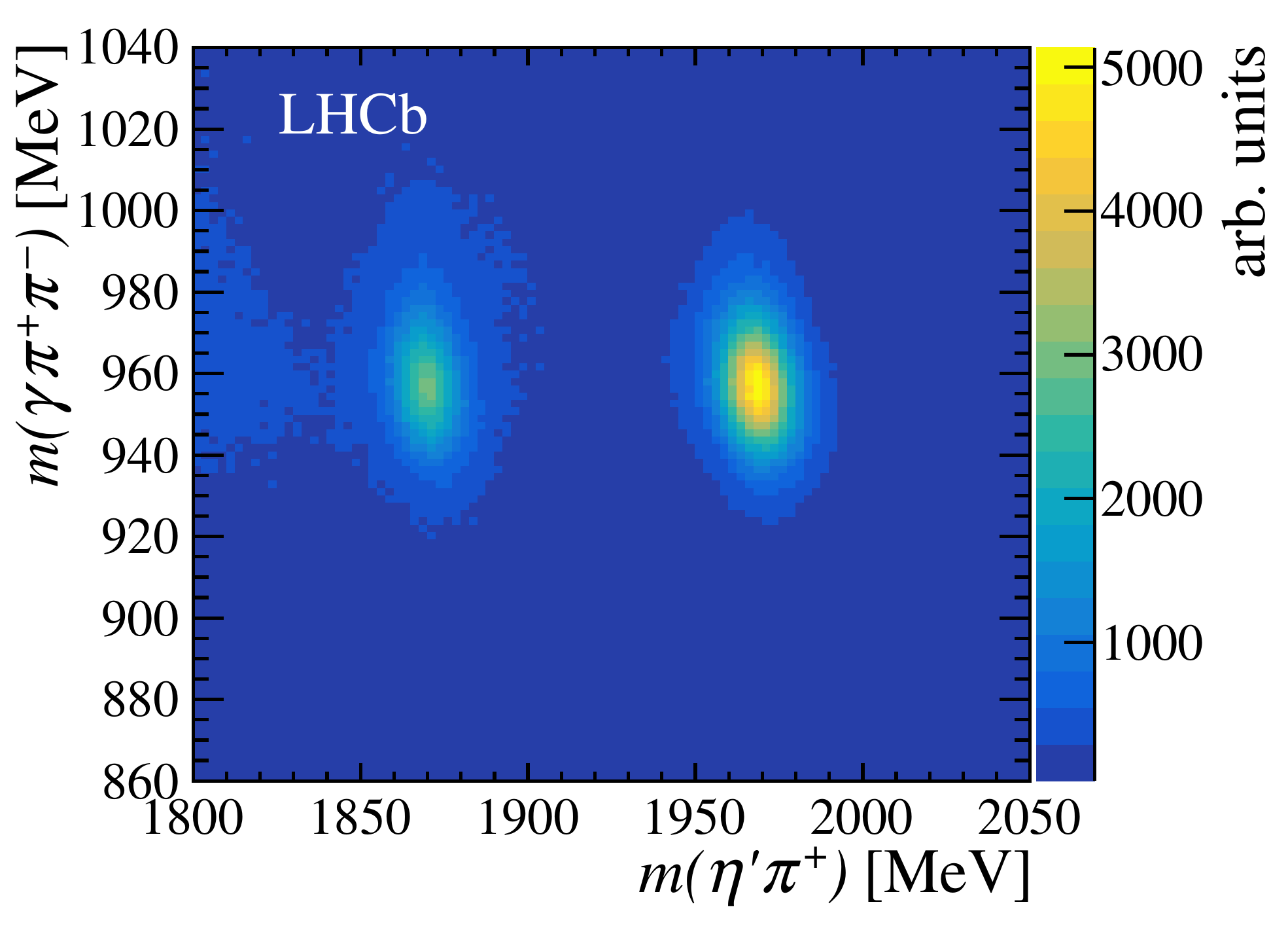}
    \includegraphics[width=.48\textwidth]{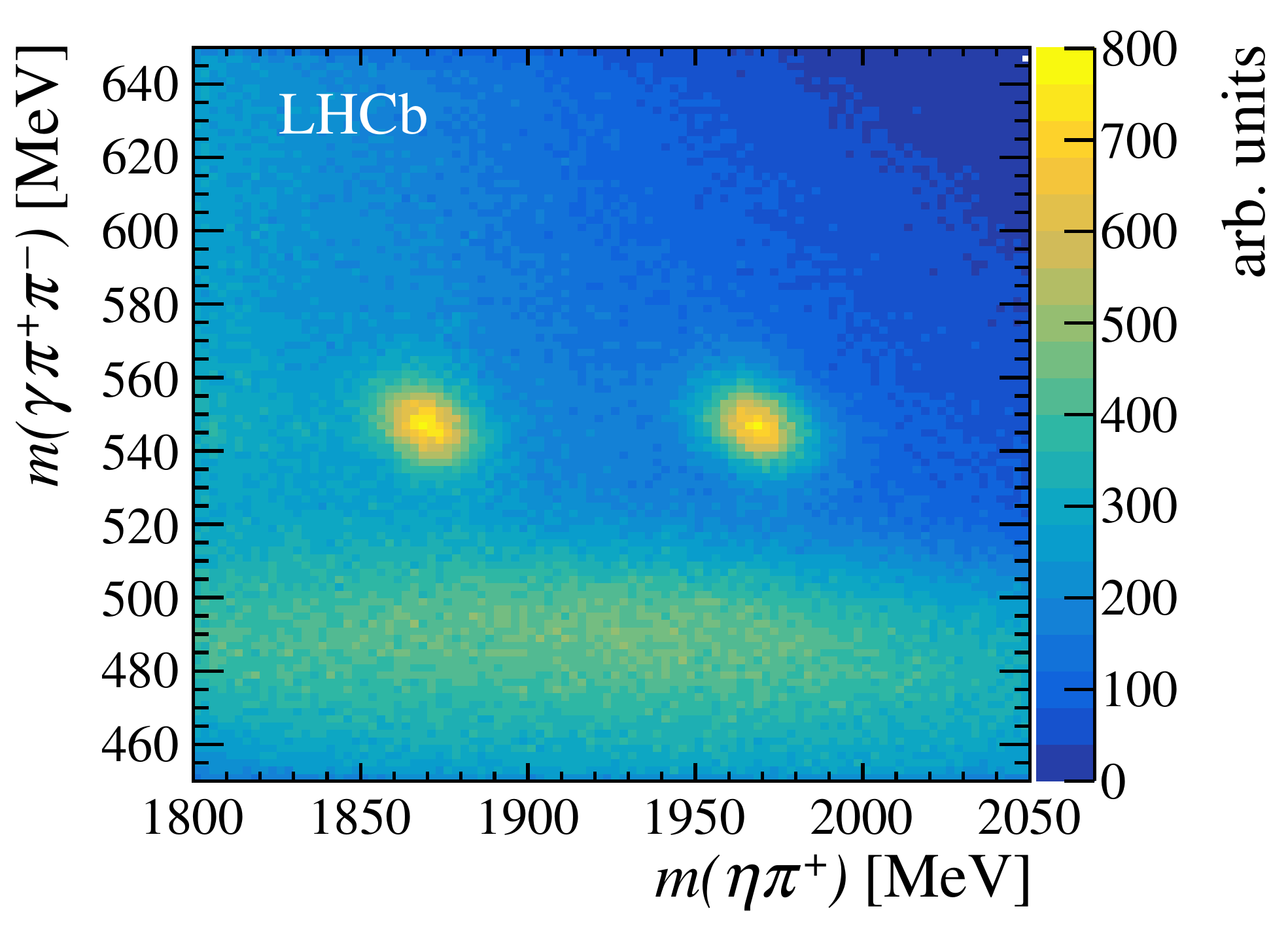}
    \caption{Two-dimensional distributions of $m(\gamma\pip\pim)$ versus $m(\etazpr\pip)$ for (left) the \etapr\pip final state and (right) the \etaz\pip final state. The background candidates at low values of $m(\gamma\pip\pim)$ for the \etaz\pip final state arise from $\etaz\to\pip\pim\piz$ decays. These distributions correspond to a subset of the signal data sample. }
    \label{fig:m2d}
\end{figure}

Another background contribution comes from the signal decay \DorDstoEtaPi, followed by $\etaz\to\pip\pim\piz$, with the \piz meson reconstructed as a photon. This decay 
is visible in the two-dimensional distribution of $m(\gamma\pip\pim)$ versus $m(\etaz\pip)$, as seen in Fig.~\ref{fig:m2d}. 
This background shows only broad peaking behaviour in $m(\etaz\pip)$ and, in any case, is removed by the selection on \mgpp.
%
It should be noted that this decay mode could be used to measure $\ACP(\DorDstoEtaPi)$ with a dedicated analysis.

Other physics background sources from partially or misreconstructed decays, such as \mbox{$\Lc\to\etazpr p$}, $\DorDs\to\etazpr\mu^+\nu_\mu$, $\DorDs\to\etazpr\pip\piz$, and $\DorDs\to\etapr\pip ~({\rm with}~\etapr\to\etaz\pip\pim$, \mbox{$\etaz\to\gamma\gamma)$}, have been studied and found to contribute negligibly in the mass range of interest. The latter two are considered when evaluating the systematic uncertainty associated to the fit model. 

\section{Determining the raw asymmetry}
\label{sec:fit}
\begin{figure}[tb]
    \centering
    \includegraphics[width=\textwidth]{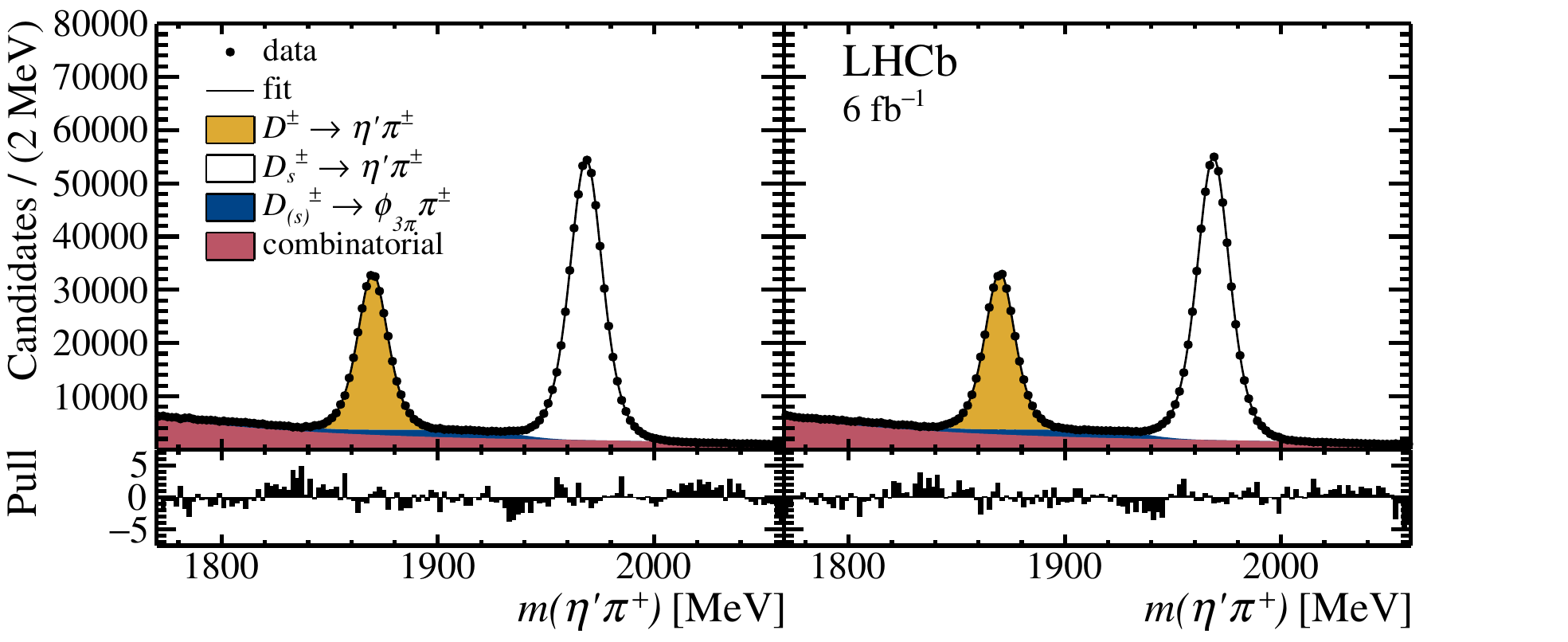}
    \caption{Distribution of the invariant mass $m(\etapr\pip)$ of (left) \DorDstoEtapPi candidates and (right) $D^{-}_{(s)}\to\etapr\pim$ candidates, summed over the eight subsamples. Candidates have \mppg in the range 936--976$\mev$. The results of the fit described in the text are superimposed. The lower panels show the normalised fit residuals (pulls).} 
    \label{fig:Mass_EtapPi}
\end{figure}

\begin{figure}[tb]
    \centering
    \includegraphics[width=\textwidth]{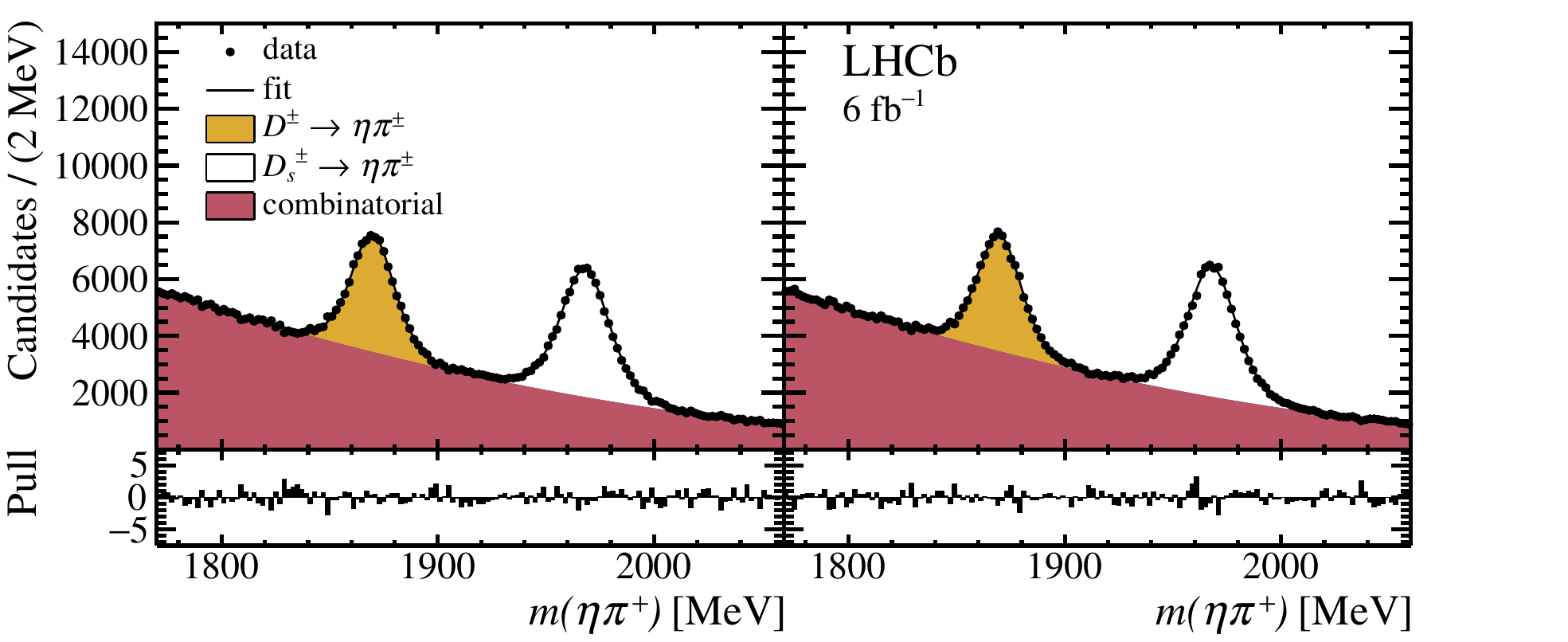}
    \caption{Distribution of the invariant mass $m(\etaz\pip)$ of (left) \DorDstoEtaPi candidates and (right) $D^{-}_{(s)}\to\etaz\pim$ candidates, summed over the eight subsamples. Candidates have \mppg in the range 526--570$\mev$. The results of the fit described in the text are superimposed. The lower panels show the normalised fit residuals (pulls).}
    \label{fig:Mass_EtaPi}
\end{figure}

\begin{figure}
    \centering
    \includegraphics[width=.48\textwidth]{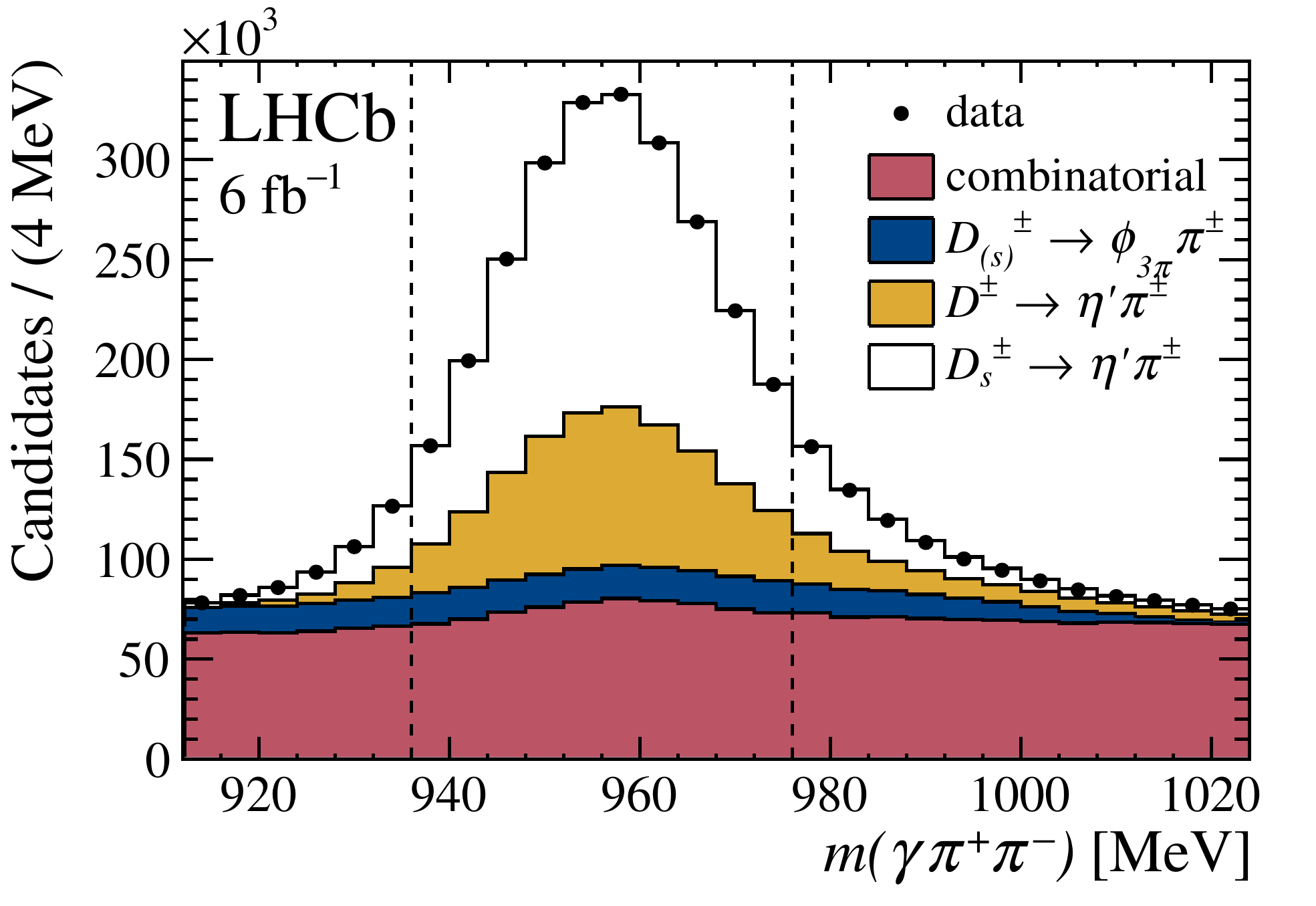}
    \includegraphics[width=.48\textwidth]{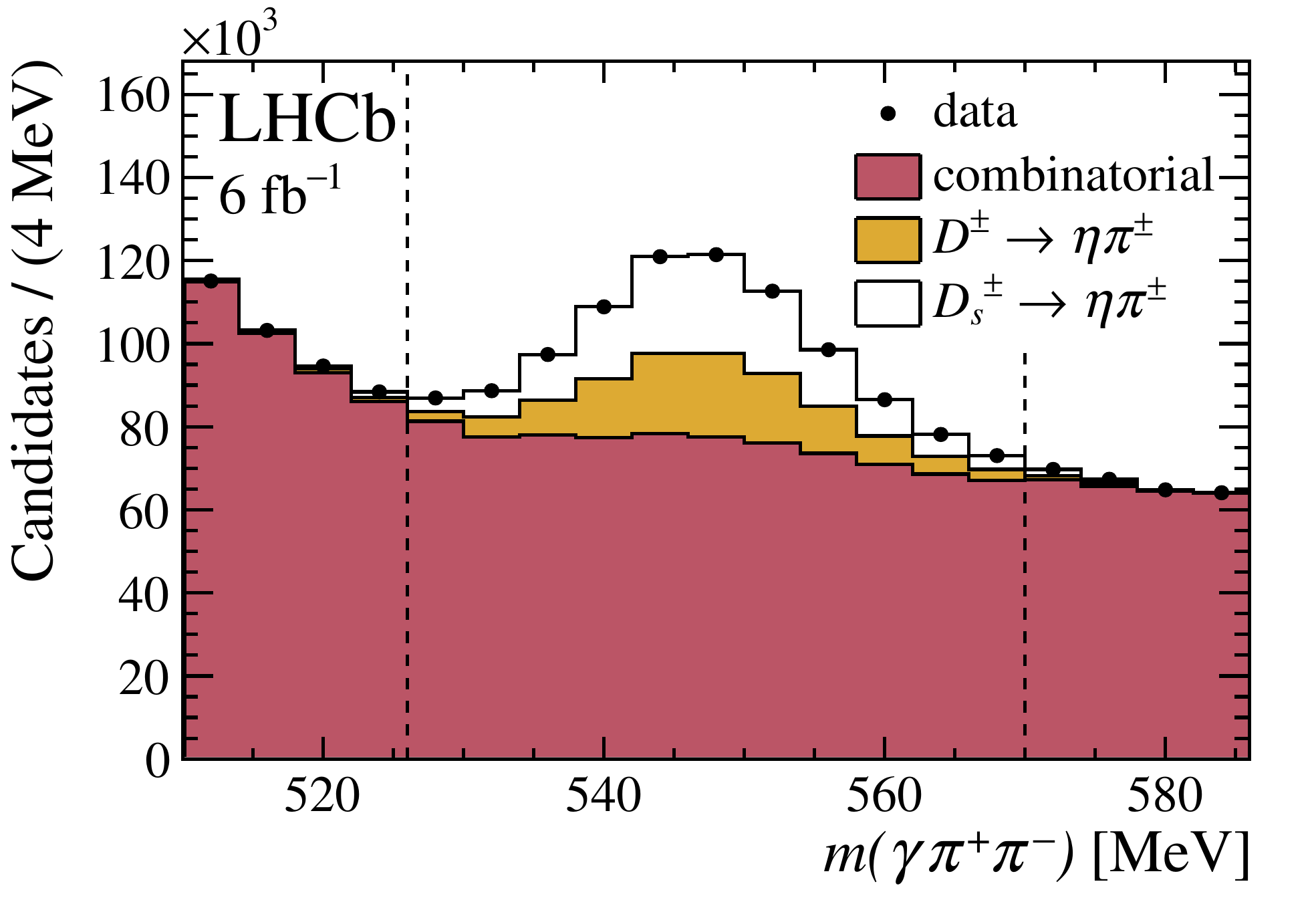}
    \caption{Distributions of the invariant mass \mppg for (left) \DorDstoEtapPi candidates and \mbox{(right) \DorDstoEtaPi candidates}, with the projections of the fit superimposed, summed over the subsamples and the \DorDs charges. In this figure, the \mppg mass ranges are enlarged with respect to the baseline fit. The default mass ranges are indicated by the vertical dashed lines.} 
    \label{fig:Mass_EtapEta}
\end{figure}

The distributions of the invariant mass $m(\etapr\pipm)$ of 
the $\DorDspm\to\etapr\pi^{\pm}$ candidates, separated by charge, 
are shown in Fig.~\ref{fig:Mass_EtapPi} and the corresponding distributions for the \mbox{$\DorDspm\to\etaz\pipm$} candidates are shown in Fig.~\ref{fig:Mass_EtaPi}. 
For each data subsample
(described in Sec.~\ref{sec:selection}),
the raw asymmetry is independently determined using a simultaneous, binned, extended maximum-likelihood fit to the $m(\etazpr\pip)$ distributions of positively and negatively charged $D^{\pm}_{(s)}$ candidates in the mass range 1770--2060 MeV. This fit to the \DorDs invariant mass is also performed simultaneously in 4 MeV intervals of the unconstrained \mgpp mass, in the mass range 936--976 \mev for the \etapr channel or 526--570 \mev for the \etaz channel. The distributions of the unconstrained \mgpp mass for \DorDstoEtapPi and \DorDstoEtaPi candidates are shown in Fig.~\ref{fig:Mass_EtapEta}.
The total charge-integrated \Ds and \Dp yields in each \mgpp bin are also determined in the fit.

The \Ds and \Dp mass peaks are each described with a Johnson SU function~\cite{ref:JohnsonSU}, while the combinatorial background component is modelled by a third-order Chebyshev polynomial. The mean of the Johnson SU function corresponding to the \Ds peak is left to vary as a function of \mgpp.
Such variation is expected due to the small correlations between these two quantities. The mean of the Johnson SU function of the \Ds peak in the $i^{\rm th}$ \mgpp interval, $\mu_\Ds^i$, is parameterised by 
\begin{equation} \label{eq:quadratic}
    \mu_\Ds^i = \mu_\Ds^p + a[M^i-M^p] + b[M^i-M^p]^2,
\end{equation}
where $M$ is shorthand for \mgpp, and the superscript $p$ refers to the value in the interval corresponding to the peak of the \mgpp distribution. The values of $\mu_\Ds^p$, $a$ and $b$ are determined in the fit.  
The fit also determines $\Delta m \equiv m(\Ds)-m(\Dp)$, which is allowed to vary linearly as a  function of \mgpp, thus determining the position of the \Dp peak in each interval of \mgpp.
Similarly, the widths of both peaks are determined by the fit and allowed to vary quadratically as a function of \mgpp. The choice of the allowed variation (linear or quadratic) is made empirically, choosing the simplest functional form that describes the data. These means and widths may be different for positively and negatively charged candidates. The two parameters describing the shape of the tails of the Johnson SU functions are shared among the \Ds and \Dp peaks, positively and negatively charged candidates and \mgpp intervals. 
The background yields and the parameters describing the combinatorial background are determined independently in intervals of \mgpp and for positively and negatively charged candidates. 

For the channels \DorDstoEtapPi a dedicated component to the fit for the background due to \DorDstoPhiPiThreePi decays is included.   
The shape of this component is obtained from simulated events. The overall yield of this background is free in the fit, while the relative contribution arising from \Dp and \Ds decays is fixed to the value estimated from known branching fractions and the relative efficiency determined from simulation.

The results of the fit, summed over the eight subsamples, for the \DorDstoEtapPi and \DorDstoEtaPi channels are shown in Figs.~\ref{fig:Mass_EtapPi} and \ref{fig:Mass_EtaPi}, respectively. The total charge-integrated signal yields for the \DstoEtapPi and \DtoEtapPi channels are
$(1\,085.7 \pm 1.2)\times 10^3$ and 
$(555.4 \pm 0.9)\times 10^3$, respectively. The obtained yields for the \DstoEtaPi and \DtoEtaPi are 
$(135.8 \pm 0.7)\times 10^3$ and $(110.8 \pm 0.7)\times 10^3$, respectively. The normalised fit residuals (residuals divided by the corresponding uncertainty) for the \DorDstoEtapPi channel show minor enhancements in correspondence with the endpoint of the of \DorDstoPhiPiThreePi peaking background (whose position is fixed in the fit) and close to the low-mass tail of the \DorDs signal component. This suggests an imperfect model for this background component, which is treated as a systematic uncertainty (Sec.~\ref{sec:systematics}). The normalised fit residuals for the \DorDstoEtaPi channel do not show a similar effect. 

The projections of the fit results on \mppg, showing the contribution determined in the fit for each component in each \mppg interval, are shown in Fig.~\ref{fig:Mass_EtapEta}. A small peaking contribution is visible in the combinatorial component, indicating the presence of correctly reconstructed \etazpr candidates paired with random companion pions. For the \DorDstoEtaPi channels, the large shoulder at low values of \mppg, attributed by the fit to the combinatorial component, is due to $\DorDs\to\etaz\pip$ decays, followed by the decay $\etaz\to\pip\pim\piz$, as discussed in Sec.~\ref{sec:backgrounds}. This figure was made using wider \mppg mass ranges than those used for the baseline fit. 
\section{Control channel analysis}
As described in Sec.~\ref{sec:Introduction}, the control channels \DorDstoPhiPi, followed by the decay $\phi\to\Kp\Km$, are used to cancel production and detection asymmetries. The triggering and selection of the candidates of the control channels follow closely those of Ref.~\cite{LHCb-PAPER-2019-002}.
In particular, kaon daughters of the \Pphi meson are required to satisfy particle-identification criteria compatible with being kaons and the sum of their \pt must be greater than 2\gev. Their invariant mass $m(K^+K^-)$ must lie in the range 1010--1030$\mev$. The pion candidate must have particle-identification information compatible with a pion and $\pt > 1 \gev$. As for the signal channels, the pion candidate must satisfy fiducial requirements that remove tracks from areas of the detector with large detector asymmetries. Each \DorDs candidate must have a large flight distance and a large flight-distance significance. Furthermore, the direction of the momentum of the \DorDs candidate must be compatible with the direction defined by its production and decay vertices. The \chisq per degree of freedom of the vertex fit for the \DorDs candidate must be small. The \DorDs candidate transverse momentum must satisfy \mbox{$\pt>2700 \mev$}. 
These channels have significantly lower background than the signal modes owing to the absence of neutral particles in the final state. 

To achieve satisfactory cancellation of production and detector asymmetries, the control sample candidates are weighted such that the distributions of kinematic and trigger variables simultaneously match the distributions observed in the signal sample. Specifically, the distributions of the transverse momentum and pseudorapidity of the companion pion and the \DorDs candidates, and of the type of trigger decision (whether independent of the final state particles or not) are matched to the signal distributions. 
The algorithm to determine the weights~\cite{Rogozhnikov:2016bdpnew} is trained on distributions from signal and control sample data candidates. The signal distributions are background-subtracted using the {\em sPlot} technique~\cite{Pivk:2004ty}, with $m(\etazpr\pip)$ as the discriminating variable; the background in the control sample is negligible. 

\begin{figure}
    \centering
    \includegraphics[width=.9\textwidth]{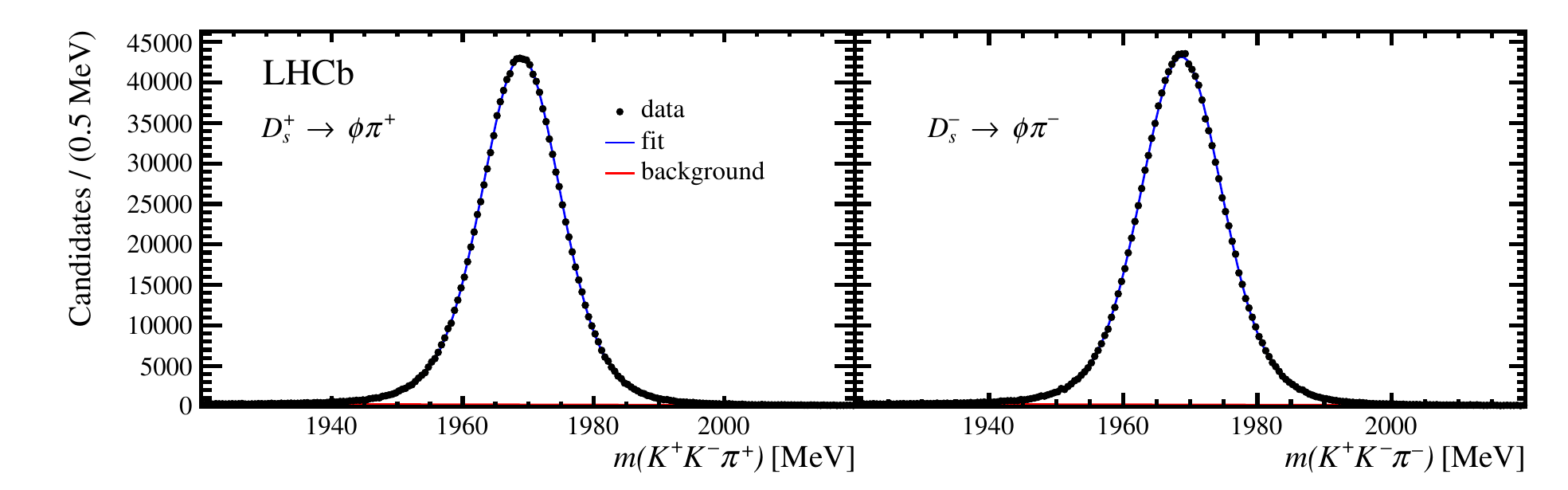}
    \includegraphics[width=.9\textwidth]{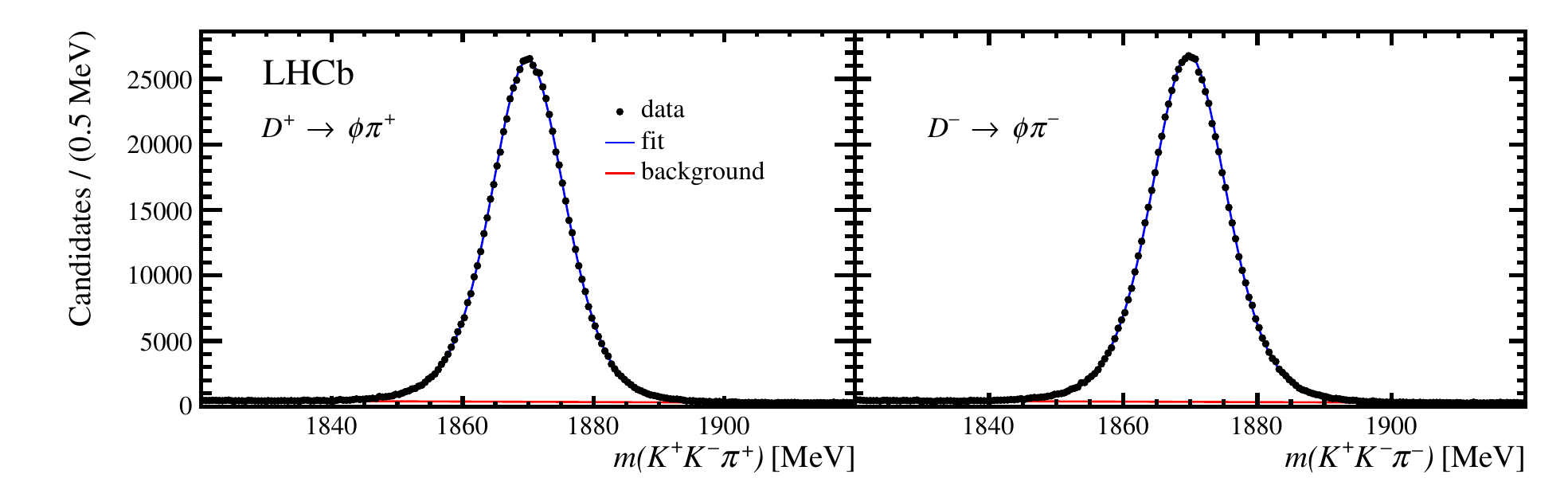}
    \caption{Mass distributions with corresponding fits of (top) $\Ds\to\phi\pip$ candidates and \mbox{(bottom) $\Dp\to\phi\pip$} candidates. The candidates are weighted to match the signal distributions for the $\etapr\pip$ channels.  The positively charged candidates are shown on the left and the negatively charged candidates on the right. The blue curves represent the total fit, while the red curves show the very small background component. The data shown are from the 2017, magnet up polarity subsample. 
    }
    \label{fig:cc_fit}
\end{figure}

The control channel data is divided into eight subsamples, as is done for the signal. The invariant mass distribution $m(\Kp\Km\pipm)$ for one subsample is shown in Fig.~\ref{fig:cc_fit}. For each subsample, the raw asymmetry is independently determined using a simultaneous binned \chisq fit to the weighted positive and negative candidates, in the range 1820--1920\mev for \Dp modes and 1920--2020\mev for \Ds modes. Each control sample channel is fit twice, employing the weighting appropriate for either the $\etapr\pip$ or the $\etaz\pip$ final state. The fit model consists of the sum of a Johnson SU function and a Gaussian function for the \DorDs peak, and a linear function for the combinatorial background.  The parameters describing the tail of the Johnson SU function, the widths of the Johnson~SU and Gaussian functions and the relative proportion of the signal described by the Gaussian component are shared among positive and negative candidates. All other fitted parameters are determined separately for positive and negative candidates. 

The result of the fit for \Araw for one control channel subsample, for both the \Ds and the \Dp decays, is shown in Fig.~\ref{fig:cc_fit}. 
The total charge-integrated signal yields for the \DstoPhiPi and \DtoPhiPi channels are  $(27\,928\pm5)\times 10^3$ and $(16\,276\pm4)\times 10^3$, respectively. \section{Systematic uncertainties}
\label{sec:systematics}

Several sources of systematic uncertainty on the \CP asymmetry measurements are presented in Table~\ref{tab:syst}.
\begin{table}[t]
    \centering
        \caption{Systematic uncertainties associated to values of \Araw (\%). }
    \label{tab:syst}
    \begin{tabular}{lcccc}
    \hline\hline
        Source &  \DstoEtapPi & \DtoEtapPi & \DstoEtaPi & \DtoEtaPi \\
    \hline
        Fit model, signal & 0.04 & 0.04 & 0.10 & 0.16\\
        Fit bias & 0.01 & 0.01 & 0.01 & 0.02 \\ 
        Secondary decays & 0.06 & 0.03 & 0.06 & 0.03 \\
        \AKK & 0.01 & 0.02 & 0.01 & 0.02 \\ 
        Fit model, control & 0.03 & 0.00 & 0.03 & 0.00 \\
        Weighting & 0.01 & 0.01 & 0.02 & 0.00 \\
    \hline
    Total & 0.08 & 0.06 &  0.12 & 0.16 \\    
    \hline\hline
    \end{tabular}
\end{table}
The  choice  of  the  fit  model  to  describe  the  signal  mass distributions is a significant source of systematic uncertainty.  This is evaluated by fitting, with the baseline model, pseudoexperiments generated according to several alternative models, the parameters of which have been previously determined from the experimental data. In particular, the background model is altered by using a  higher-order polynomial function to represent the combinatorial component, including additional background components for partially or misreconstructed decays, or varying their parameterisation, while a sum of two Gaussian functions is used as an alternative parameterisation for the signal. An estimate of the uncertainty caused by possible bias introduced by the fitting procedure is investigated using pseudoexperiments. 
A significant contribution to the systematic uncertainty also arises from secondary decays, where the \DorDs is produced in the decay of a \bquark hadron. Such decays have a slightly different value of \Aprod, causing a shift in the measured value of \ACP. To estimate this contribution, the fraction of secondary background in the signal is derived from simulation, while the production asymmetry of secondary decays is determined on data using a control sample enhanced in such decays. 

Sources of systematic uncertainties associated to the analysis of the control channels are also included.
 The uncertainty due to the fit model of the control channels is determined, as for the signal channels, by considering alternative fit models.  Detection asymmetries due to a momentum asymmetry between the kaons in the $\phi\to\Kp\Km$ decay have a small associated uncertainty~\cite{LHCb-PAPER-2019-002}. The systematic uncertainty associated to the weighting procedure is evaluated by repeating the fits to the control channels without performing the weighting, and the resulting shift in \Araw is taken as the systematic uncertainty.  

The systematic uncertainties described here were determined for the full data set, \ie not separately for the different subsamples, and hence are applicable to the final \ACP values given in the next section. The systematic uncertainties for the signal channels and the control channels are added in quadrature. 

\section{Results}

The \CP asymmetry is obtained by subtracting the value of \Araw of the control channel from that of the signal channel. For the case of \Dp decays, which are Cabibbo-suppressed, the previously measured value of $\ACP(\Dp\to\phi\pip)$ is taken into account, while for Cabibbo-favoured \Ds decays, $\ACP(\DstoPhiPi) = 0$ is assumed, resulting in
\begin{align*}
    \ACP(\Ds\to\etazpr\pi) & =  \Araw(\Ds\to\etazpr\pi) - \Araw(\Ds\to\phi\pip), \\
    \ACP(\Dp\to\etazpr\pi) & =  \Araw(\Dp\to\etazpr\pi) - \Araw(\Dp\to\phi\pip) + \ACP(\Dp\to\phi\pip),
\end{align*}
where $\ACP(\Dp\to\phi\pip) = (0.005 \pm 0.051)\%$~\cite{LHCb-PAPER-2019-002}.

\begin{figure}
    \centering
    \includegraphics[width=\textwidth]{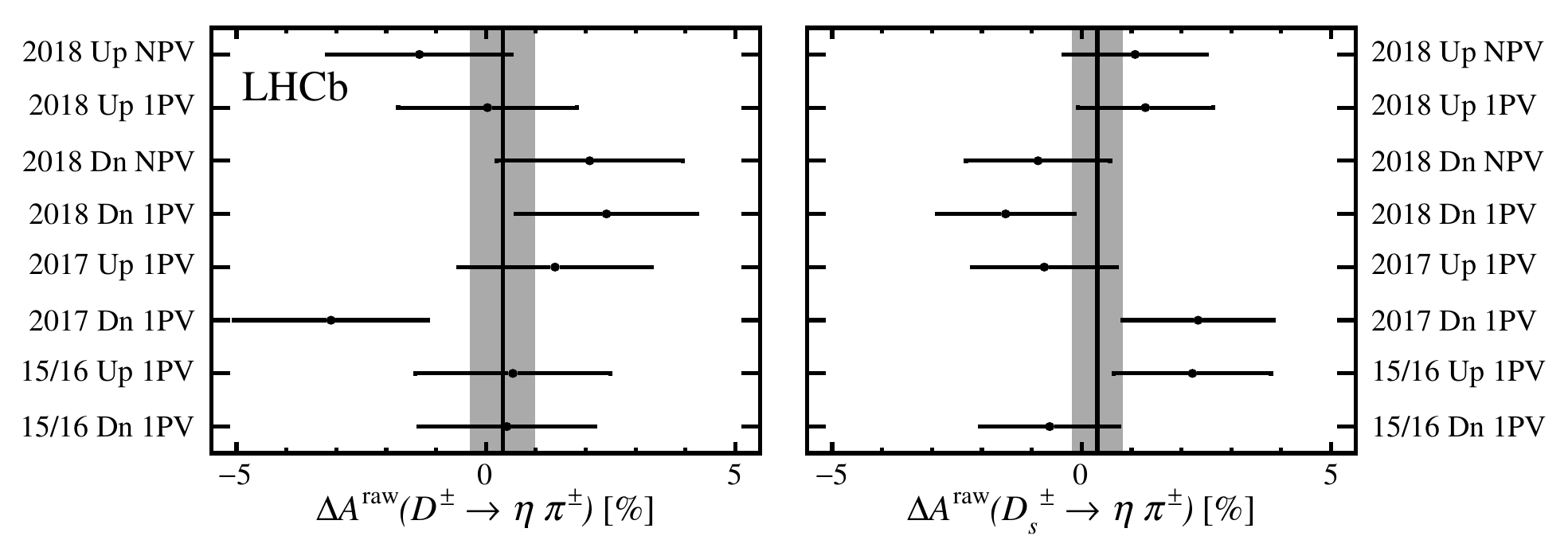}
    \includegraphics[width=\textwidth]{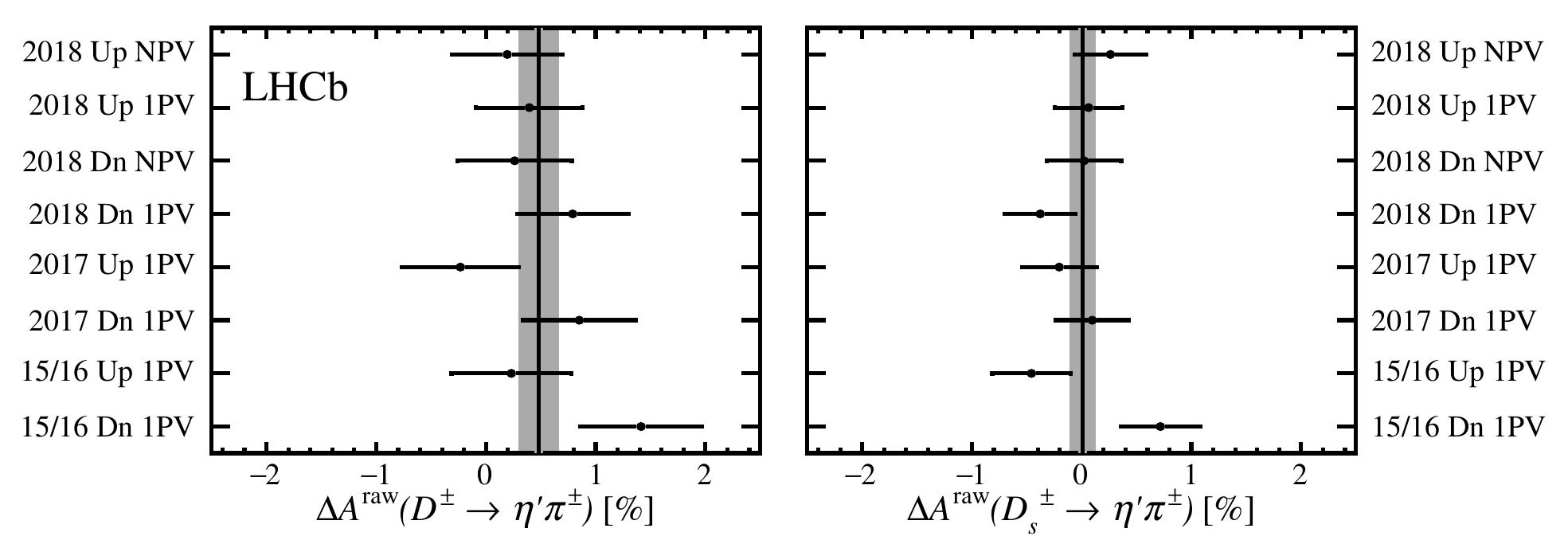}
    \caption{Measured values of $\Delta\Araw$ for the individual subsamples for (upper left) the \DtoEtaPi channel, (upper right) the \DstoEtaPi channel, (lower left) the \DtoEtapPi channel and (lower right) the \DstoEtapPi channel. The subsample labels indicate year of data taking (15/16 = 2015 + 2016), magnet polarity (``Up'' or ``Dn'') and number of PVs (``1PV'' for one primary vertex and ``NPV'' for $N_{\rm PV}>1$). The vertical lines and the grey bands indicate the weighted averages and the corresponding statistical uncertainties.}
    \label{fig:subsample_acp_Eta}
\end{figure}

The results for $\Delta\Araw = \Araw(\DorDstoEtazpPi) - \Araw(\DorDstoPhiPi)$
for the separate subsamples are shown in Fig.~\ref{fig:subsample_acp_Eta}.
The value of \ACP is determined independently for each of the eight subsamples and a weighted sum is performed to obtain the overall value of \ACP, yielding
\begin{align*}
      \ACP(\DtoEtaPi) & = (0.34 \pm 0.66 \pm 0.16 \pm 0.05)\%, \\
      \ACP(\DstoEtaPi) & = (0.32 \pm 0.51 \pm 0.12)\%, \\
      \ACP(\DtoEtapPi) & = (0.49 \pm 0.18 \pm 0.06 \pm 0.05)\%, \\
      \ACP(\DstoEtapPi) & = (0.01 \pm 0.12 \pm 0.08)\%, 
\end{align*}
where the first uncertainty is statistical, the second is  systematic and the third, relevant for the \Dp channels, is due to the uncertainty on $\ACP(\Dp\to\phi\pip)$. The use of a single control channel for two signal channels, \eg, \DstoPhiPi for \DtoEtaPi and \DtoEtapPi, introduces a small correlation between the corresponding \ACP measurements. This correlation is found to be less than 1\%.
  
These results are combined with previous \lhcb measurements based on an independent data sample~\cite{LHCb-PAPER-2016-041} or a different \etaz decay channel~\cite{LHCb-PAPER-2021-001}. A weighted sum, with weights determined using only the statistical uncertainties, is performed. The systematic uncertainties are uncorrelated between previous and current measurements. The results of the combination are
\begin{align*}
      \ACP(\DtoEtaPi) & = \phantom{+}(0.13 \pm 0.50 \pm 0.18 )\%, \\
      \ACP(\DstoEtaPi) & = \phantom{+}(0.48 \pm 0.42 \pm 0.17)\%, \\ 
      \ACP(\DtoEtapPi) & = \phantom{+}(0.43 \pm 0.17 \pm 0.10)\%,  \\       
      \ACP(\DstoEtapPi) & = (-0.04 \pm 0.11 \pm 0.09)\%, 
\end{align*}
where the first uncertainty is statistical and the second uncertainty is systematic, which includes the uncertainty on the externally input values of \ACP of the control channels. 

In summary, searches for \CP violation in the decays \DorDstoEtaPi and \DorDstoEtapPi are performed using $pp$ collisions, collected by the \lhcb experiment at centre-of-mass energy $\sqrt{s}=13$ TeV, corresponding to 6 \invfb of integrated luminosity. The results are consistent with the absence of \CP violation in these decay modes and with previous measurements~\cite{Won:2011ku,Belle:2021ygw,LHCb-PAPER-2016-041,LHCb-PAPER-2021-001}. The measurements are the most precise to date for the \DtoEtaPi, \DtoEtapPi and \DstoEtapPi channels.


\section*{Acknowledgements}
%
%
\noindent We express our gratitude to our colleagues in the CERN
accelerator departments for the excellent performance of the LHC. We
thank the technical and administrative staff at the LHCb
institutes.
We acknowledge support from CERN and from the national agencies:
CAPES, CNPq, FAPERJ and FINEP (Brazil); 
MOST and NSFC (China); 
CNRS/IN2P3 (France); 
BMBF, DFG and MPG (Germany); 
INFN (Italy); 
NWO (Netherlands); 
MNiSW and NCN (Poland); 
MCID/IFA (Romania); 
MICINN (Spain); 
SNSF and SER (Switzerland); 
NASU (Ukraine); 
STFC (United Kingdom); 
DOE NP and NSF (USA).
We acknowledge the computing resources that are provided by CERN, IN2P3
(France), KIT and DESY (Germany), INFN (Italy), SURF (Netherlands),
PIC (Spain), GridPP (United Kingdom), 
CSCS (Switzerland), IFIN-HH (Romania), CBPF (Brazil),
Polish WLCG  (Poland) and NERSC (USA).
We are indebted to the communities behind the multiple open-source
software packages on which we depend.
Individual groups or members have received support from
ARC and ARDC (Australia);
Minciencias (Colombia);
AvH Foundation (Germany);
EPLANET, Marie Sk\l{}odowska-Curie Actions and ERC (European Union);
A*MIDEX, ANR, IPhU and Labex P2IO, and R\'{e}gion Auvergne-Rh\^{o}ne-Alpes (France);
Key Research Program of Frontier Sciences of CAS, CAS PIFI, CAS CCEPP, 
Fundamental Research Funds for the Central Universities, 
and Sci. \& Tech. Program of Guangzhou (China);
GVA, XuntaGal, GENCAT and Prog.~Atracci\'on Talento, CM (Spain);
SRC (Sweden);
the Leverhulme Trust, the Royal Society
 and UKRI (United Kingdom).

\addcontentsline{toc}{section}{References}
\bibliographystyle{LHCb}
\typeout{} 
\bibliography{main,standard,LHCb-PAPER,LHCb-CONF,LHCb-DP,LHCb-TDR}
 
\newpage
\centerline
{\large\bf LHCb collaboration}
\begin
{flushleft}
\small
R.~Aaij$^{32}$\lhcborcid{0000-0003-0533-1952},
A.S.W.~Abdelmotteleb$^{50}$\lhcborcid{0000-0001-7905-0542},
C.~Abellan~Beteta$^{44}$,
F.~Abudin{\'e}n$^{50}$\lhcborcid{0000-0002-6737-3528},
T.~Ackernley$^{54}$\lhcborcid{0000-0002-5951-3498},
B.~Adeva$^{40}$\lhcborcid{0000-0001-9756-3712},
M.~Adinolfi$^{48}$\lhcborcid{0000-0002-1326-1264},
H.~Afsharnia$^{9}$,
C.~Agapopoulou$^{13}$\lhcborcid{0000-0002-2368-0147},
C.A.~Aidala$^{77}$\lhcborcid{0000-0001-9540-4988}
% ...

{\footnotesize \it

% ...
$^{9}$Universit{\'e} Clermont Auvergne, CNRS/IN2P3, LPC, Clermont-Ferrand, France\\
% ...
$^{13}$LPNHE, Sorbonne Universit{\'e}, Paris Diderot Sorbonne Paris Cit{\'e}, CNRS/IN2P3, Paris, France\\
% ...
$^{32}$Nikhef National Institute for Subatomic Physics, Amsterdam, Netherlands\\
% ...
$^{40}$Instituto Galego de F{\'\i}sica de Altas Enerx{\'\i}as (IGFAE), Universidade de Santiago de Compostela, Santiago de Compostela, Spain\\
% ...
$^{44}$Physik-Institut, Universit{\"a}t Z{\"u}rich, Z{\"u}rich, Switzerland\\
% ...
$^{48}$H.H. Wills Physics Laboratory, University of Bristol, Bristol, United Kingdom\\
% ...
$^{50}$Department of Physics, University of Warwick, Coventry, United Kingdom\\
% ...
$^{54}$Oliver Lodge Laboratory, University of Liverpool, Liverpool, United Kingdom\\
% ...
$^{62}$Syracuse University, Syracuse, NY, United States\\
% ...
$^{77}$University of Michigan, Ann Arbor, MI, United States, associated to $^{62}$\\
\bigskip
% ...
}
\end{flushleft}

\end{document}